# Spline Based Intrusion Detection in Vehicular Ad Hoc Networks (VANET)

David A. Schmidt, Mohamad S. Khan, Brian T. Bennett

*Department of Computing, East Tennessee State University*
*Roy S. Nicks Hall, 365 Stout Dr., Johnson City, TN 37604*

schmidtda@etsu.edu
khanms@mail.etsu.edu
bennetbt@etsu.edu

*Abstract*— Intrusion detection systems (IDSs) play a crucial role in the identification and mitigation for attacks on host systems. Of these systems, vehicular ad hoc networks (VANETs) are particularly difficult to protect due to the dynamic nature of their clients and their necessity for constant interaction with their respective cyber-physical systems. Currently, there is a need for a VANET-specific IDS that can satisfy these requirements. Spline function-based IDSs have shown to be effective in traditional network settings. By examining the various construction of splines and testing their robustness, the viability for a spline-based IDS can be determined.

Keywords- Intrusion Detection, Spline, Vehicular Ad Hoc Networks, Machine Learning, Internet of Things, Internet of Vehicles

## I. INTRODUCTION

The internet of things (IoT) has become an area of interest as society begins to connect a vast number of computer systems to the internet. Due to this societal switch and consequent volume of information becoming digitalized, intrusion detection has become a high-priority concern. With the implementation of so many types of computer systems, attempts have been made to create an optimal intrusion detection system (IDS) for each domain in IoT. Among these domains, the internet of vehicles (IoV), consisting of vehicular ad hoc networks (VANETs), has proven to be an especially difficult domain due to the dynamic nature of their clients and the complexity of the criteria associated with its optimal IDS. Currently, the use of spline functions within IDSs has been pioneered in multiple domains with various levels of success. The purpose of this article is to examine the role of an IDS, the obstacles associated with VANET, and define a preliminary spline-based IDS configuration for a VANET.

## II. PURPOSE OF AN INTRUSION DETECTION SYSTEM

An Intrusion Detection System (IDS) is an application that identifies attacks against a host system. IDSs identify these attacks by processing system-centric events using various machine learning and data mining techniques. Ideally, these techniques allow an IDS to function efficiently and effectively.

The effectiveness of an IDS stems from its ability to distinguish between normal processing and attacks, the speed with which it identifies attacks, and how well it determines an attack's type. At minimum, an IDS should distinguish between denial of service attacks, probes, unauthorized elevation of privilege, and remote access attacks. Efficiency results from lowering the time required for processing and the resources required for these identifications to occur. Overall, an optimal IDS should balance efficiency and effectiveness to remain robust.

### A. Probing

Sharma and Kaul state "The most common type of attack, a probing is an attack that monitors a target system to collect data and identify weaknesses" [1]. If an attacker can exploit a weakness, they can use this information to compromise the integrity of a system. A myriad of probing attacks exists. Often, these attacks involve the exploitation of a system's hardware, such as an open-access port. More often, probing attacks are the result of a user's incompetence and sensitive information is divulged to the attacker.

### B. Denial of Service (DoS) Attacks

Denial of Service (DoS) is a type of attack where an attacker overburdens system resources [1]. This prevents users from making legitimate requests from one of the compromised resources, thus, denying access. DoS typically occurs when an attacker abuses a feature of a system by exploiting bugs or poor designs within the system. "Often, these types of attacks are classified based on the resource that is compromised" [1]. For instance, a UPD socket flood is a DoS attack that floods a target with packets. This gridlocks a system because it is unable to process every request.

### C. User to Root (U2R) Attacks

"User to Root (U2R) attacks are a type of attack where an intruder attempts to gain root access to a target system" [1]. Once an attacker gains root access, they can obtain administrator privileges, thus compromising the system and the integrity of its contents. This type of attack usually occurs in conjunction with a buffer-overflow exploit but can also be found in attacks such as code injection techniques.

### D. Remote to User (R2U) Attacks

"Remote to User (R2U) attacks are a type of attack where an attacker exploits a system over a network by sending malicious packets in order to expose the target system to vulnerabilities" [1]. The attacker then exploits the target system to gain user access and exploit vulnerabilities as a local user. Typically, this type of attack occurs as phishing; however, this attack may also occur if an attack alters networking control protocols.

### III. Obstacles Associated with Vehicular Ad Hoc Networks

Unlike traditional networks which consist of simple static client-server relationships, VANET's relationships have an increased complexity. This is due to the dynamic nature of their clients, the multifaceted nature of their cyber-physical systems, and the safety-critical nature of the environments in which they are implemented. Within the system, fast-moving autonomous vehicles act as clients who, in turn, interact with a cyber-physical system consisting of cellular service towers, roadside units, and other clients within range. Development of an optimal IDS will not only need to account for the client's safety in terms of the five types of attacks, but also the integrity of the cyber-physical system. Outside the realm of cyber-attacks, damage to power-grid infrastructure and roadside units via physical attacks will need to be accounted for as well for a VANET IDS [2]. Any damage received by the cyber-physical system may lead to inconsistencies in communication. If these inconsistencies are not corrected, the resulting communication breakdown may result in a breach of security.

Due to the need for constant communication between a vehicle and the cyber-physical system, network latency is a high priority concern. There are many components that must be accounted for to provide a fast, responsive connection. However, the limiting factor for an optimal connection is in essence how fast a client can detect an attack and the speed in which it communicates to the corresponding system to mitigate any damage [3]. In a situation where a vehicle is traveling at high speeds or even at lower speeds in densely populated areas, the health and safety of individuals are at stake. This safety-criticality must not be taken lightly. Calculated decisions must be made within fractions of a second, always with preservation of life as a top priority. Failure to meet this criterion is a failure to meet the standards of an optimal IDS.

### IV. Splines

A spline is a mathematical depiction of a continuous function consisting of points, called knots, that allows the user to manipulate the shape of curves [4]. A user interfaces with a spline function by entering a specific number of knots. A curve is then created between each of these knots. A curve that passes through one of these knots is deemed an interpolating curve and one that passes near them is deemed an approximation curve [5]. Due this ability, spline functions are useful when shaping two-dimensional and, in the case of B-Splines, three-dimensional depictions of functions up to the complexity of a cubic polynomial [6]. Any computation in a higher complexity environment can result in a loss of accuracy due to a decrease in the amount of interpolation corresponding with an increase in approximation [7]. Based on their continuity, splines can be organized into different subsets including, but not limited to, linear, quadratic, cubic, and basis splines.

#### A. Linear Splines

A linear spline is the simplest form of interpolation. It is constructed piece-wise from linear functions creating two–point interpolating polynomials. An example of a linear interpolating spline (Fig. 1) and an equation (1) representing a linear spline interpolation, as seen in [8], can be seen below.

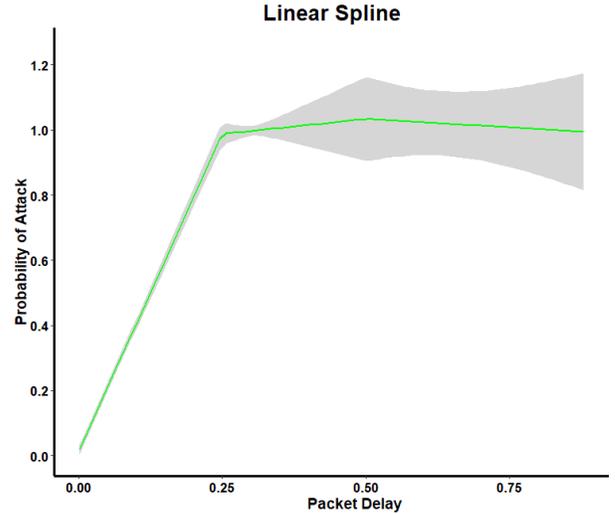

Fig. 1 Visual representation of a linear spline. Grey denotes three degrees of freedom. Knot placement occurs at the 0.25, 0.50, and 0.75 packet delay quantile.

A linear spline $g$ is given by

$$g(x) = \sum_{i=i}^{m} y_i B_{i,i}(x)$$

And satisfies the interpolation conditions

$$g(x_i) = y_i, \text{ for } i = 1, \ldots, m - 1 \text{ and}$$
$$\lim_{x \to x_m^-} g(x) = y_m \quad (1)[8]$$

*1) Linear Spline Interpretation:* Due to its two-point polynomial interpolation, linear splines are confined to a strict degree of freedom when interpolating data. These degrees of freedom are derived from the variable placement of knots during the spline's construction which, in turn, are confined by its interpolation conditions [4]. As demonstrated in Fig. 1, these conditions allow polynomials only of the second degree to be utilized, thus restricting the possible spline variations. This variation, although minimal, is key to the spline's robustness when classifying data as it allows for placement of the best fit linear spline.

#### B. Quadratic Splines

Increasing in complexity, quadratic spline construction is like that of linear splines. However, rather than consisting of piece-wise linear functions, it is constructed from piece-wise quadratic functions. An example of a quadratic spline (Fig. 2) and an equation (2) representing its piece-wise construction, as demonstrated in [9], can be seen below.

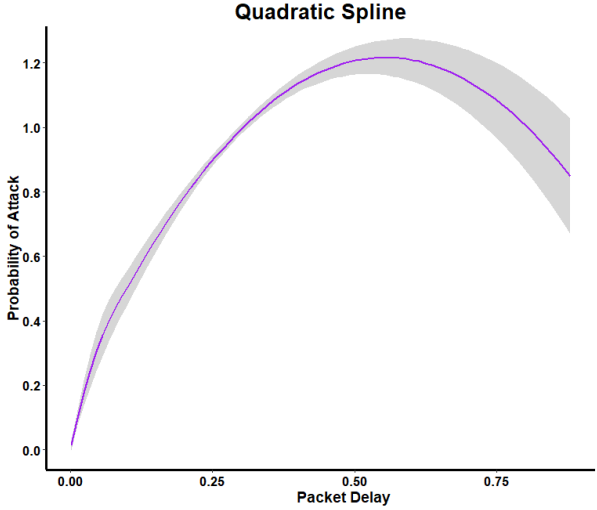

Fig. 2 Visual representation of a quadratic spline. Grey shading denotes three degrees of freedom. Knot placement occurs at the 0.25, 0.50, and 0.75 packet delay quantile.

The quadratic spline $S_{2,2}(x)$ is constructed as:

$$P_1(x) = a_1 + b_1 x + c_1 x^2, \quad \text{on} \quad [-1, 0]$$
$$P_2(x) = a_1 + b_1 x + c_1 x^2, \quad \text{on} \quad [0, 1]$$

And $S_{2,2}(x)$ interpolates the given data points,

$$P_1(-1) = a_1 + b_1 + c_1 = 0$$
$$P_1(0) = a_1 = 1$$
$$P_1(0) = a_2 = 1$$
$$P_1(1) = a_2 + b_2 + c_2 = 3$$

The quadratic spline function is given as

$$S_{2,2}(x) = \begin{cases} 1 + 2x + x^2, & \text{on } [-1, 0] \\ 1 + 2x, & \text{on } [-1, 0] \end{cases} \quad (2)[9]$$

*1) Quadratic Spline Interpretation:* Like linear splines, quadratic splines also possess a varying degree of freedom. However, due to the nature of their construction, quadratic splines can interpolate knots at a second degree which increases the possible interpolant variations for the spline [4]. As seen in Fig. 2, the quadratic spline contains a much wider area of coverage than the linear spline with identical degrees of freedom. Based on the placement of its knots, the increased freedom of the quadratic spline can specify the way it interpolates knots, thus increasing its robustness when classifying data.

C. Cubic Splines

The highest level of complexity cubic spline construction is like that of linear splines, but rather than consisting of piece-wise linear functions, it is constructed from piece-wise cubic functions. An example of a cubic spline (Fig. 3) and an equation (3) representing a quadratic spline's construction, as seen in [9], can be seen below.

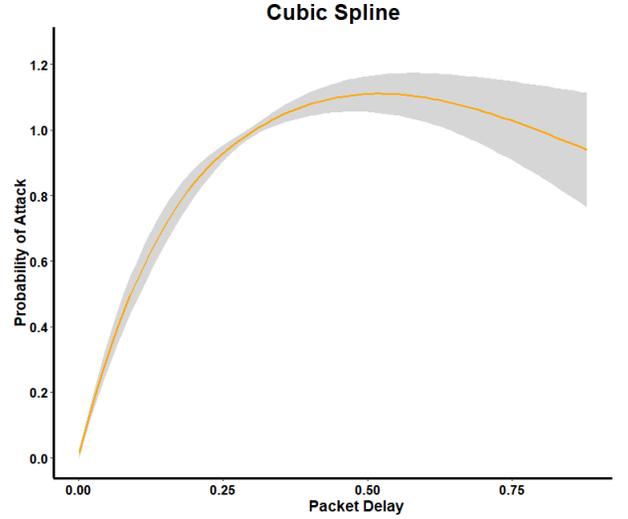

Fig. 3 Visual representation of a cubic spline. In this instance. Grey shading denotes three degrees of freedom. Knot placement occurs at the 0.25, 0.50, and 0.75 packet delay quantile.

Assuming the data used for construction is $\{(x_i, f_i)\}_{i=0}^n$, and $S_{3,n}(x)$ is a cubic spline constructed as:

$$P_1(x) = a_1 + b_1 x + c_1 x^2 + d_1 x^3, \quad x \in [x_0, x_1],$$
$$P_2(x) = a_2 + b_2 x + c_2 x^2 + d_2 x^3, \quad x \in [x_1, x_2],$$
$$\vdots$$
$$P_n(x) = a_n + b_n x + c_n x^2 + d_n x^3, \quad x \in [x_{n-1}, x_n];$$

which interpolates the given data points

$$S_{3,n}(x) = f_i, \quad i = 0, 1, \dots, n$$

the cubic spline function is given as

$$S_{3,n}(x) = \begin{cases} P_1(x) = a_1 + b_1 x + c_1 x^2 + d_1 x^3, & x \in [x_0, x_1] \\ P_2(x) = a_2 + b_2 x + c_2 x^2 + d_2 x^3, & x \in [x_1, x_2], \\ \vdots \\ P_n(x) = a_n + b_n x + c_n x^2 + d_n x^3, & x \in [x_{n-1}, x_n] \end{cases} \quad (3)[9]$$

*1) Cubic Spline Interpretation:* In comparison to the previously mentioned splines, cubic splines demonstrate the highest level of variation within the same degree of freedom. This is due to the spline's piecewise cubic construction which allows interpolation between knots to occur at the third degree, further increasing the number of possible variations for the spline [4]. With this expanded variation, cubic splines can further manipulate the manner in which they interpolate knots and further increasing its robustness when classifying data.

D. Basis Splines (B-splines)

A B-spline is a special case of spline function in which a spline of order *n* in a piece-wise constructed function of the degree *n – 1* in terms of a variable *x* [5]. If a B-spline of this order is equivalent among all knots, all possible spline functions for the set of polynomials pertaining to the B-spline can be constructed using a combination of linear B-splines with only a single unique combination for each spline [5]. An equation (4) depicting a mathematical representation of a B-spline and its construction, as demonstrated in [10], can be seen below.

A B-spline curve *P(t)* is defined by

$$P(t) = \sum_{i=0}^{n} P_i N_{i,k}(t)$$

Where:

- the control points are $\{P_i : i = 0, 1, \dots, n\}$,
- $k$ is the order of polynomial segments within the curve,
- the $N_{i,k}(t)$ are the normalized B-spline blending functions described by the order $k$ and by a non-decreasing order of real numbers

$$\{t_i : i = 0, \dots, n + k\}.$$

The $N_{i,k}$ component functions are

$$N_{\_}(i, 1)(t) \begin{cases} 1 \text{ if } u \in [t_i, t_{i+1}), \\ 0 \text{ otherwise} \end{cases}$$

Where, if $k > 1$,

$$N_{i,k}(t) = \frac{t - t_i}{t_{i+k-i} - t_i} N_{i,k-1}(t) + \frac{t_{i+k} - t}{t_{i+k} - t_{i+1}} N_{i+1,k-1}(t) \ (4)[9].$$

## V. CONCEPT FOR VANET SPLINE-BASED IDS

Although the concept of a spline-based IDSs has been implemented using traditional client-to-sever relationships [6] [7], there are limited cases of its use in IoT and few, if any, utilized within the realm of IoV. Recently, experimentation within the domain of IoV conducted by Shams, Rizaner, and Ulusoy [11] has provided a novel approach for intrusion detection using a support vector machine (SVM) in combination with a trust value table (TVT). Due to the success of this approach, the implementation of splines to this framework may further optimize the IDS. This optimization stems from the nature of the SVM where a hyper-plane is created at an optimal distance between adjacent data points [12]. It is speculated that the addition of splines may further optimize this process, increasing the accuracy when identifying malicious attacks.

## VI. PRELIMINARY EXPERIMENTATION

In order to demonstrate the viability of a spline-based IDS in an IoV environment, several spline regressions, as well as a logistic regression, were implemented on an IoV dataset provided by Shams et al. [11] (Fig 4). The data consisted of network traffic between fifty-two simulated autonomous vehicles that monitored packet delay, the number of packets dropped, and the frequency of transfer interval in both congested and non-congested environments [11]. Of these data, packet delay was selected as the independent variable due to its ability to perform as an attack predictor. Six hundred observations were randomly selected and split into subsets with an 80:20 ratio for training and testing. Confusion-matrix analysis for the splines demonstrated accuracies greater than 95%, with the B-spline holding an accuracy of 98.30% like that of the 99.17% accuracy of the logistic regression (Table 1).

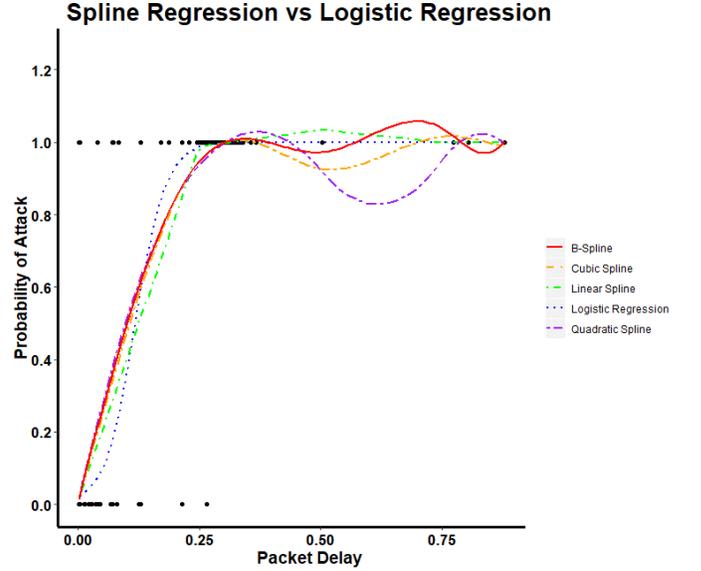

Fig. 4 Prediction of attack, indiscriminate of type, using various splines and logistic regression. Knot placement occurs at the 0.25, 0.50, and 0.75 packet delay quantile.

## VII. CONCLUSION

IDSs designed for vehicular ad hoc networks must be able to compensate for the dynamic nature of their clients, their associated safety criticality, and the multifaceted nature of their cyber-physical systems in order to be effective. Due to their ability to interpolate curves, their high levels of efficiency, and their success in traditional network environments, spline functions may prove to be vital component for an optimal VANET IDS. The use of B-splines and their ability to form unique combinations of linear functions are of extreme interest as they may provide a robust solution to mediate these obstacles. In conjunction with the ensemble of SVM and TVT, other techniques such as association and clustering analysis may also provide valuable insight for the construction of an optimal IDS, as they may allow the grouping of specific types of attacks, leading to faster classification and, therefore, more robust decision making.

## VIII. FUTURE RESEARCH

It is important to note that in this preliminary study, only the functionality of splines was discussed. Important topics such as processing speed, attack type classification, dynamic knot placement, and safety criticality for VANETs will be explored as the scope of research expands. In future research machine learning will be applied to splines for the purpose of creating a robust classifier. In turn, this classifier will be tested among variety of VANET environments that include suburban, urban, and highway situations.

Table 1. Confusion matrix analysis for logistic and spline regressions, where N denotes the number of observations.

| N = 120 | True Positive | False Positive | True Negative | False Negative | Prediction Accuracy |
|---|---|---|---|---|---|
| Logistic Regression | 61 | 1 | 58 | 0 | 99.17% |
| Linear Spline | 59 | 3 | 56 | 2 | 95.83% |
| Quadratic Spline | 59 | 1 | 59 | 1 | 98.30% |
| Cubic Spline | 59 | 2 | 59 | 0 | 98.30% |
| B-Spline | 60 | 2 | 58 | 0 | 98.30% |